\documentclass[twocolumn]{aastex61}
\pdfoutput=1 
\usepackage{amsmath,amstext}
\usepackage[T1]{fontenc}
\usepackage{apjfonts} 
\usepackage{comment}
\usepackage[figure,figure*]{hypcap}
\usepackage{multirow}
\usepackage{amsmath}
\usepackage{nccmath}

\newcommand{\lkh}{LkH$\alpha$~101}
\newcommand{\lkht}{LkH$\alpha$~101~VLA~J043017.90+351510.0}
\newcommand{\lkhc}{LkH$\alpha$~101~VLA~J043019.14+351745.6}
\newcommand{\lkhs}{LkH$\alpha$~101~VLA~J043001.15+351724.6}
\newcommand{\msec}[2]{$#1\mbox{$''\mskip-7.6mu.\,$}#2$}

\shorttitle{The distance to the LkH$\alpha$~101 cluster}
\shortauthors{Dzib et al.}

\begin{document}

\title{VLBA Determination of the Distance to Nearby Star-forming Regions. VIII. \\The LkH$\alpha$~101 cluster}

\correspondingauthor{Sergio A. dzib}
\email{sdzib@mpifr-bonn.mpg.de}

\author[0000-0001-6010-6200]{Sergio A. Dzib}
\affiliation{Max-Planck-Institut f\"ur Radioastronomie, Auf dem H\"ugel 69,
 D-53121 Bonn, Germany}
\author{Gisela N.~Ortiz-Le\'on}
\affiliation{Max-Planck-Institut f\"ur Radioastronomie, Auf dem H\"ugel 69,
 D-53121 Bonn, Germany}
\affiliation{Humboldt Fellow}
\author{L. Loinard}
\affiliation{Instituto de Radioastronom\'{\i}a y Astrof\'{\i}sica, Universidad Nacional Aut\'onoma de M\'exico, Morelia 58089, Mexico}
\affiliation{ Instituto de  Astronom\'{\i}a, Universidad Nacional Aut\'onoma de M\'exico, Apartado Postal 70-264, CdMx C.P. 04510, Mexico}
\author{A.~J.~Mioduszewski}
\affiliation{National Radio Astronomy Observatory, P.O. Box 0, Socorro, NM 87801, USA}
\author{L. F. Rodr\'{\i}guez}
\affiliation{Instituto de Radioastronom\'{\i}a y Astrof\'{\i}sica, Universidad Nacional Aut\'onoma de M\'exico, Morelia 58089, Mexico}
\author{S.-N. X. Medina}
\affiliation{Max-Planck-Institut f\"ur Radioastronomie, Auf dem H\"ugel 69,
 D-53121 Bonn, Germany}
\author{R.~M.~Torres}
\affiliation{Centro Universitario de Tonal\'a, Universidad de Guadalajara, Avenida Nuevo Perif\'erico No. 555, Ejido Dan Jos\'e Tatepozco, C.P. 48525, Tonal\'a, Jalisco, M\'exico.}

\begin{abstract}
The \lkh\ cluster takes its name from its more massive member, the \lkh\ star,
which is an $\sim15$~M$_\odot$ star whose true nature is 
still unknown. The distance to the
\lkh\ cluster has been controversial for the last few
decades, with estimated values ranging from
160 to 800 pc. We have observed members and candidate 
members of the \lkh\ cluster with signs of magnetic activity, using 
the Very Long Baseline Array, in order to measure their 
trigonometric parallax and, thus, obtain a direct measurement of their distances. 
A young star member, \lkh\ VLA J043001.15+351724.6, was
detected at four epochs as a single radio source. 
The best fit to its displacement on the plane of the sky 
yields a distance of 
535$\pm$29 pc. We argue that this is the distance
to the \lkh\,~cluster.
\end{abstract}


\nopagebreak
\section{Introduction}\label{sec:intro}

LkH$\alpha$~101 is a massive star, $M \sim15$~M$_\odot$, with an extinction of 
$A_{\rm V}\simeq10$.
It illuminates the reflection nebula NGC~1579 and has a directly imaged disk 
\cite[see the review by][and references therein]{andrews2008}. It also hosts a small 
H{\small II} region that is sustained by the 
ionized winds from its disk \citep{thum2013}.  \lkh\ is associated with a cluster of young low-mass
stars (hereafter, the \lkh\ cluster), some of which are magnetically active \citep{bw1988,stine1998,osten2009}. These 
properties strongly suggest that LkH$\alpha$~101 is a young high-mass star. However, there is an absence 
of stellar absorption features and, thus, there is no classification for its
photosphere \citep{herbig2004}. In fact, its spectroscopic properties have been compared to some 
post-main-sequence massive stars. 
As has been discussed by \cite{andrews2008}, the true nature of LkH$\alpha$~101 is
still a mystery and this problem is compounded by its large distance uncertainty.

An extended discussion of the different suggested distances to the \lkh\ cluster 
(which includes the \lkh\ star) has been presented in the review 
by \citet{andrews2008} and we summarize it here. Initially, \citet{1971ApJ...169..537H} 
estimated a distance of 
800 pc based on UBV photometry of two nearby early B-type stars.  Later,  
\citet{stine1998} suggested an entirely different value of 160 pc, arguing that the radio 
luminosities of T Tauri stars in the \lkh\ cluster would be incompatible with that of T Tauri 
stars in Taurus-Aurigae if the cluster were at 800 pc. However, \citet{herbig2004}
noted that this method is inadvisable. \citet{tuthill2002} favored a value of
d$\,\simeq340$~pc, from model constraints on the star-disk mass for \lkh\ and the proper
motions of a companion. \citet{herbig2004} obtained spectral parallax 
measurements to 40 young \lkh\ cluster members with a wide range of spectral types and estimated
d$\,\simeq700\pm200$~pc.  The conclusion of the discussion by \citet{andrews2008}
was that most of the observational constrains suggest a distance between 500~pc to 700~pc. 
These authors also noted that the two different methods to identify cluster membership, by
\citet{feigelson1999} and \citet{feigelson2005}, are in good agreement when applied to the
\lkh\ cluster if its distance is about 550 pc.  Clearly, the past suggested distances to the cluster have very
large uncertainties because they are based only on interpretations of the properties 
of the stars. To date, there are no direct 
measurements of distances to any of the members of the \lkh\ cluster and the most common 
assumed distance is 700~pc. An accurate distance to this region is fundamental to
constrain the true nature of the \lkh\,~star.

Magnetically active young stars are excellent targets to measure trigonometric parallax 
and, thus, determine direct distances. They can be observed with 
the Very Long Baseline Interferometry (VLBI) technique \cite[see][for recent results]{dzib2016,ol2017serp,
ol2017ophi,kounkel2017}. By employing this technique, we observed
suspected magnetically active young stars in the \lkh\ cluster to measure their distance and
provide a more accurate distance to the cluster.

\section{Observations and Data Calibration}

We observed six young star members and a candidate member of the \lkh\ cluster. 
These target sources are listed in Table~\ref{tab:targets}. We
used the Very Long Baseline Array \cite[VLBA;][]{napier94}, operated 
by the Long Baseline Observatory (LBO), under projects BD165 and BD207. 
The observations used the multi-phase center capability provided by 
the VLBA DifX digital correlator \citep{2011PASP..123..275D}. The 
observations of the first project were carried out on 2012 October 9 and October 11, 
at a wavelength of 3.6~cm ($\nu$ = 8.42 {\rm GHz}). 
In the first session (October 9), the first four targets 
in Table~\ref{tab:targets} were observed, while the other three were observed 
as part of the second session (October 11).  
Following the successful detections of three target sources, 
we initiated a series of multi-epoch observations (project BD207) at a wavelength of 6.0~cm 
($\nu$ = 4.5 {\rm GHz}) starting in 2017 March and, subsequently, we obtained
a new observation every three months. 
The change of receiver was due to three reasons. First, the primary beam size is
larger and we can cover all seven sources in a single pointing.  Second,  the 
new 6.0 cm band receiver is more sensitive than the 3.6 cm receiver. Finally, because 
of their negative spectral indices, the target sources are brighter at longer wavelengths. 
These last two reasons increase the SNR of the detection, increasing the precision of
the position measurement. 

\begin{table}[!hb]
\small
\centering
\renewcommand{\arraystretch}{1.1}
\caption{Observed sources. The names and infrared classes from \cite{osten2009}.}
\begin{tabular}{lcccc}\hline\hline
Name & IR & \\
(\lkh\ VLA) & Class & Detected?\\
\hline
J043010.87+351922.4&III&No\\
J043016.04+351726.9&III&No\\
J043017.90+351510.0\tablenotemark{a}&...&Yes\\
J043019.14+351745.6&II&Yes\\
J042953.98+351848.2&III&No\\
J043001.15+351724.6\tablenotemark{b}&III&Yes\\
J043002.64+351514.9&II&No\\
\hline\hline
\label{tab:targets}
\end{tabular}
\tablenotetext{a}{Candidate member.}
\tablenotetext{b}{This paper is based on the detections of this star.}
\end{table}

\begin{table*}[!htbp]
\small
\centering
\renewcommand{\arraystretch}{1.1}
\caption{Observation dates, synthesized beam size, and noise levels of the final maps around LkH$\alpha$ 101 VLA J043001.15+351724.6, as well as measured source position and flux densities.}
\begin{tabular}{lcccccccccc}\hline\hline
Mean UT date &Julian Day & Synthesized beam & $\sigma_{\rm noise}$& $\alpha$(J2000.0) & $\sigma_{\alpha}$ & $\delta$(J2000.0) & $\sigma_\delta$ & $f_\nu$ \\
(yyyy.mm.dd/hh:mm)& &  ($\theta_{\rm maj}\times\theta_{\rm min};$ P.A.) & ($\mu$Jy$\,$bm$^{-1}$)& $04^{\rm h}30^{{\rm m}}$& & $35^{\circ}17'$& & (mJy)\\
\hline\noalign{\smallskip}
2012.10.09/10:17&2456211.93&\msec{0}{0021}$\times$\msec{0}{0008};\,6.5$^\circ$&36&1\rlap.{$^{\rm s}$}146281 &0\rlap.{$^{\rm s}$}000006& 24\rlap.{''}43957&0\rlap.{''}00014&0.26$\pm$0.04 \\
2017.03.25/21:46&2457838.41&\msec{0}{0036}$\times$\msec{0}{0012};\,18.6$^\circ$&23&1\rlap.{$^{\rm s}$}146702 &0\rlap.{$^{\rm s}$}000008& 24\rlap.{''}41308&0\rlap.{''}00017&0.20$\pm$0.02 \\
2017.06.17/16:16&2457922.18&\msec{0}{0034}$\times$\msec{0}{0012};\,15.4$^\circ$&21&1\rlap.{$^{\rm s}$}146934 &0\rlap.{$^{\rm s}$}000007& 24\rlap.{''}41223&0\rlap.{''}00022&0.16$\pm$0.02 \\
2017.09.11/10:38&2458007.94&\msec{0}{0046}$\times$\msec{0}{0015};\,-1.0$^\circ$&27&1\rlap.{$^{\rm s}$}147056 &0\rlap.{$^{\rm s}$}000002& 24\rlap.{''}41143&0\rlap.{''}00007&0.81$\pm$0.05 \\
\hline\hline
\label{tab:hri}
\end{tabular}
\end{table*}

\begin{figure}[!ht]
   \centering
  \includegraphics[height=0.35\textwidth,trim= 0 0 0 0, clip]{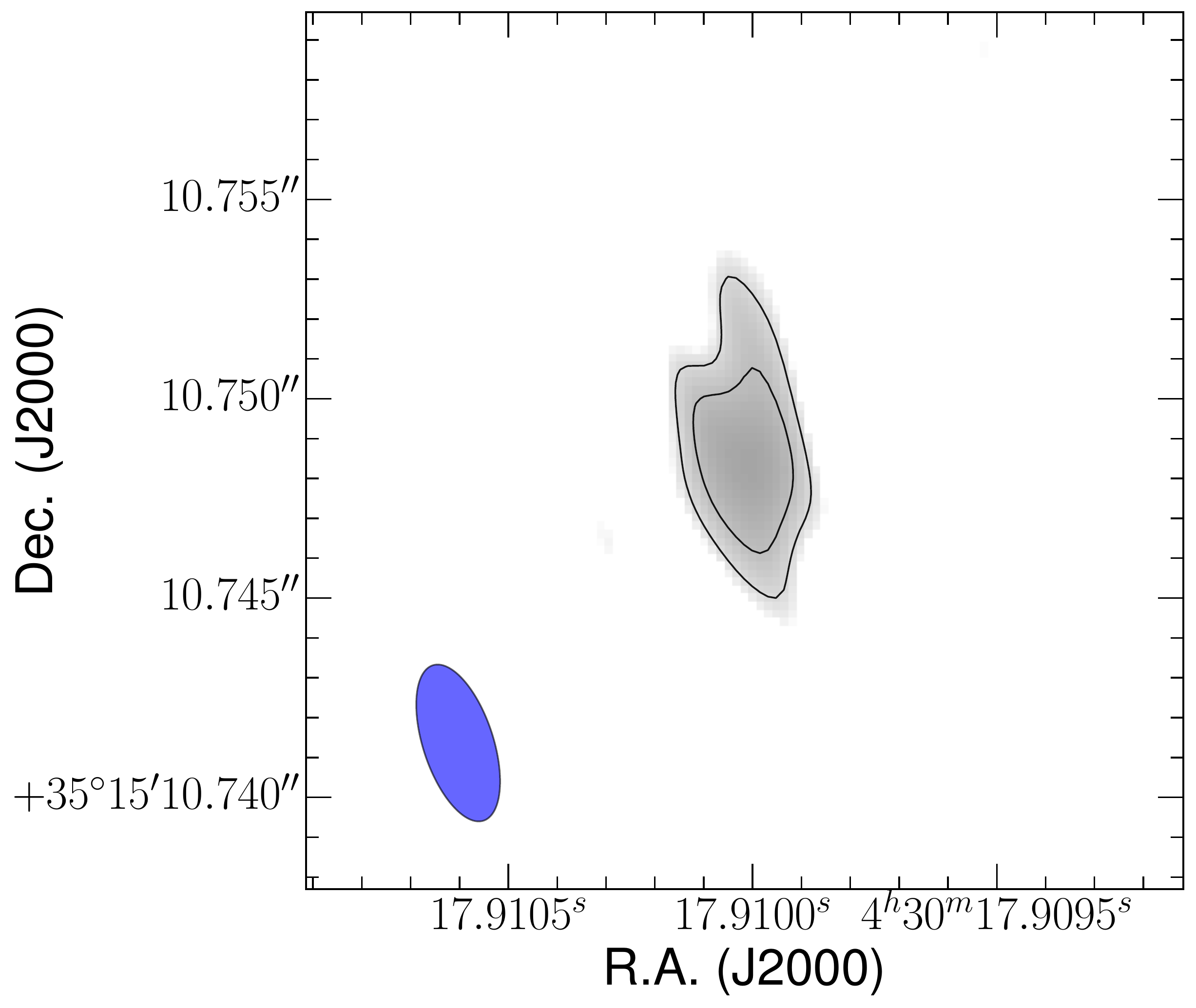}  
  \includegraphics[height=0.35\textwidth,trim= 0 0 0 0, clip]{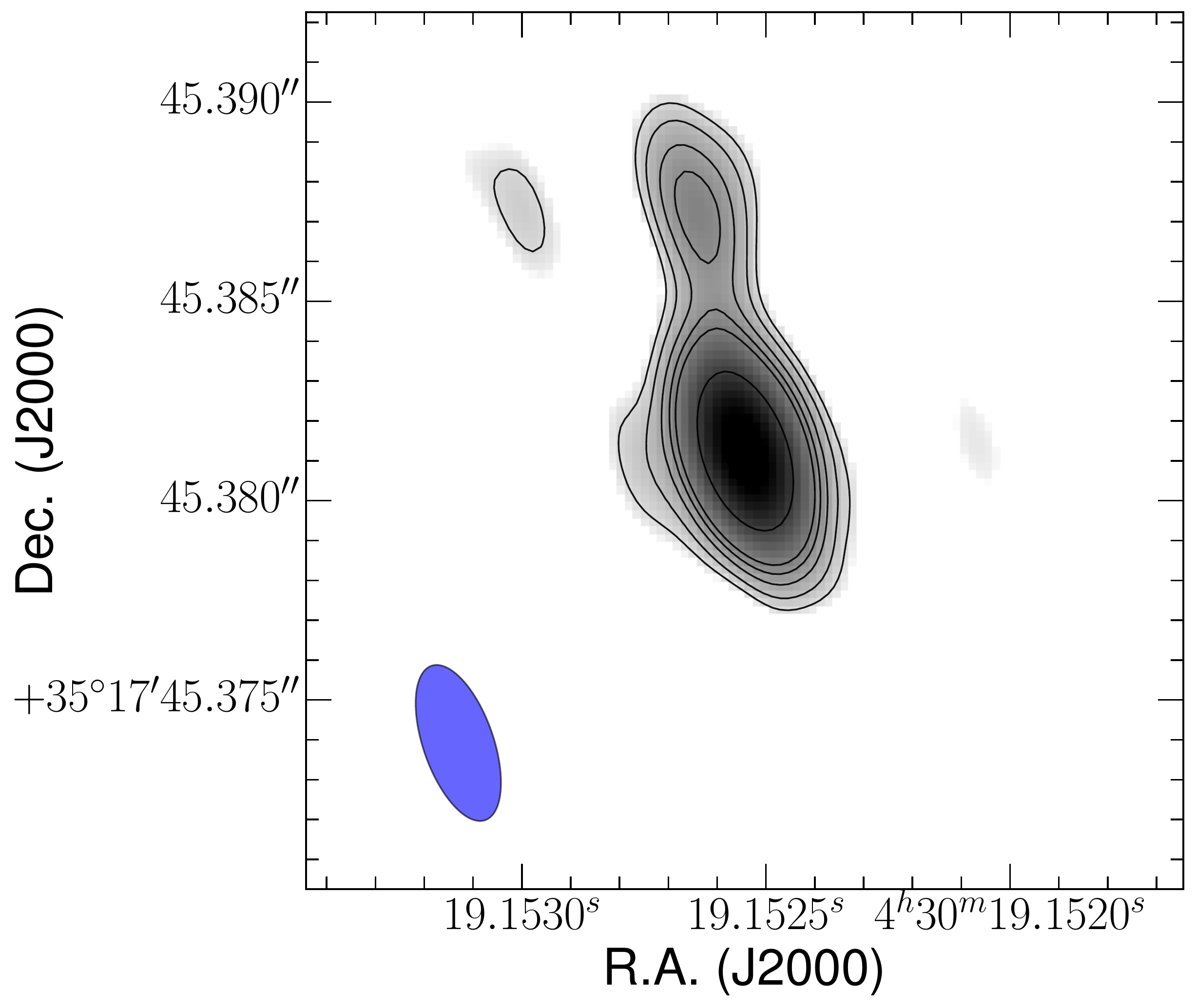}  
\caption{\lkht\ (top) and \lkhc\ (bottom) as detected on 2017 March 25. The noise levels
are 31 $\mu$Jy beam$^{-1}$ and 38~$\mu$Jy beam$^{-1}$ for the top and bottom images, respectively.
The contour levels are -3, 4, 6, 9, 12, 15, 30, and 50 times the noise level. The 
size of the synthesized primary beam for both images is 4.1 mas $\times$ 1.6 mas; 
P.A.= 18\rlap{.}$^\circ$5, and is displayed as a filled blue ellipse in the bottom left corner of each image.}
   \label{fig:so}
\end{figure}

\begin{figure*}[!ht]
\begin{tabular}{cc}
   \centering
  \includegraphics[height=0.35\textwidth,trim= 0 0 40 0, clip]{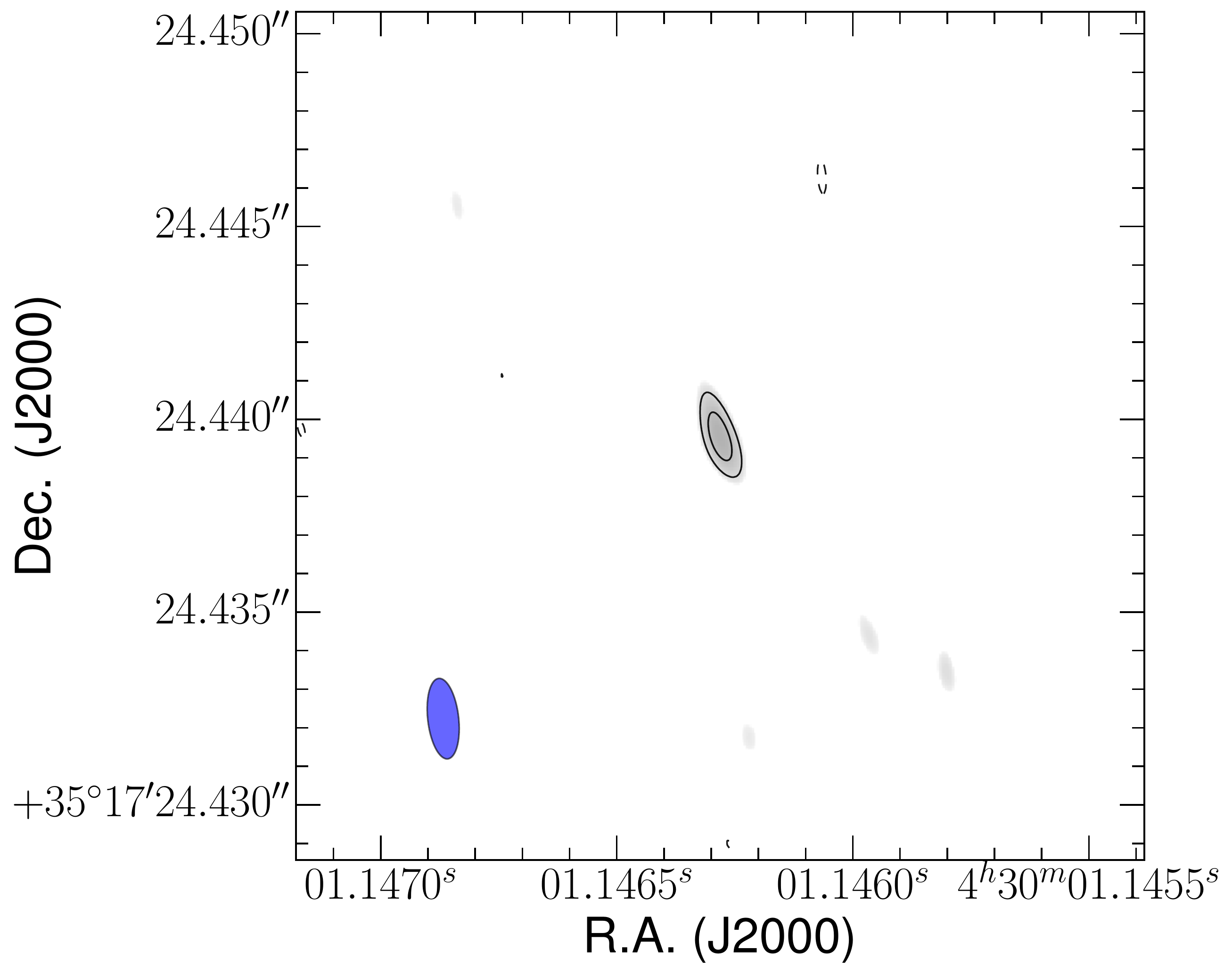}
  \put(-72,160){2012.10.09}&
    \includegraphics[height=0.35\textwidth,trim= 0 0 0 0, clip]{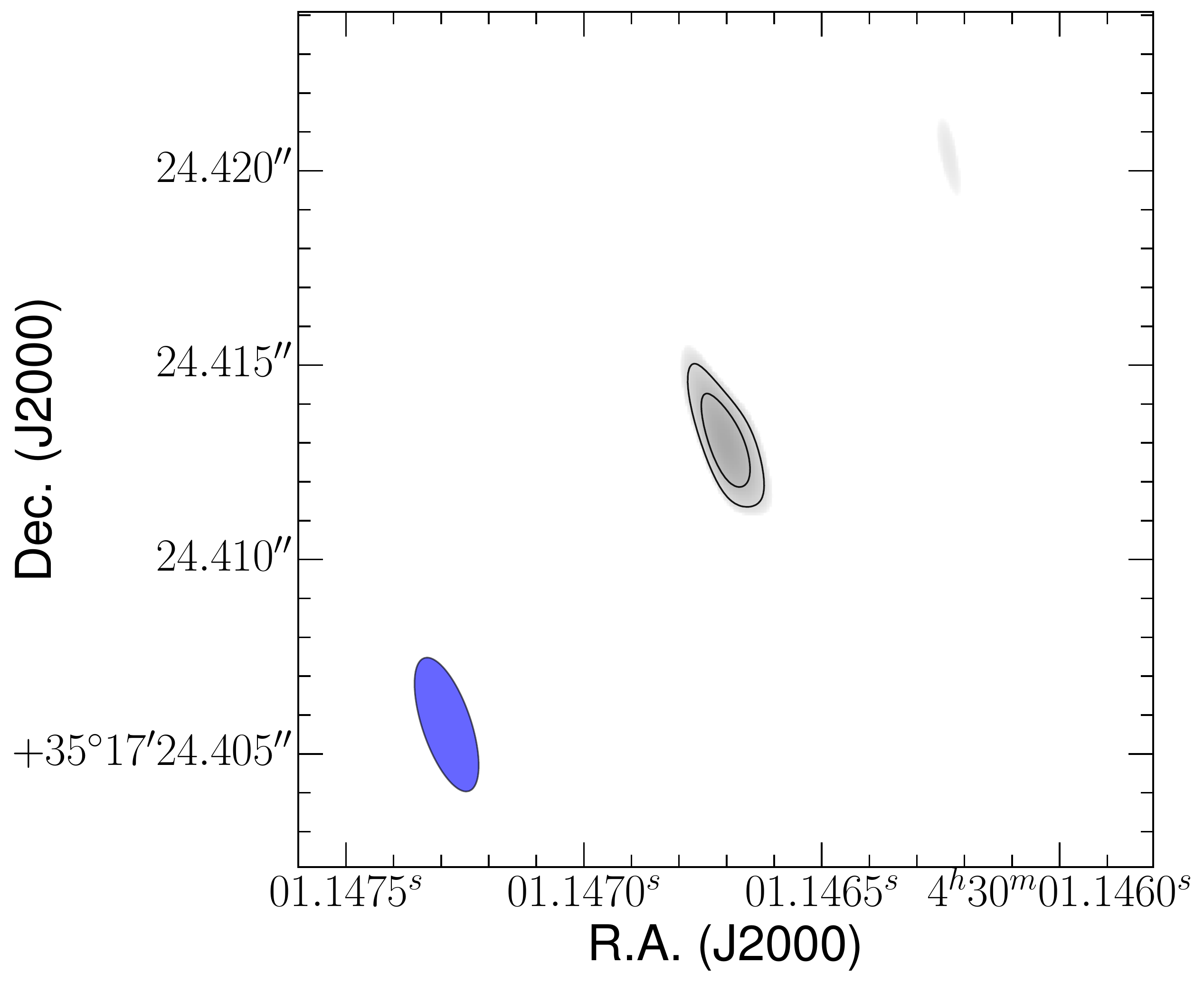} 
    \put(-72,160){2017.03.25}\\
  \includegraphics[height=0.35\textwidth,trim= 0 0 0 0, clip]{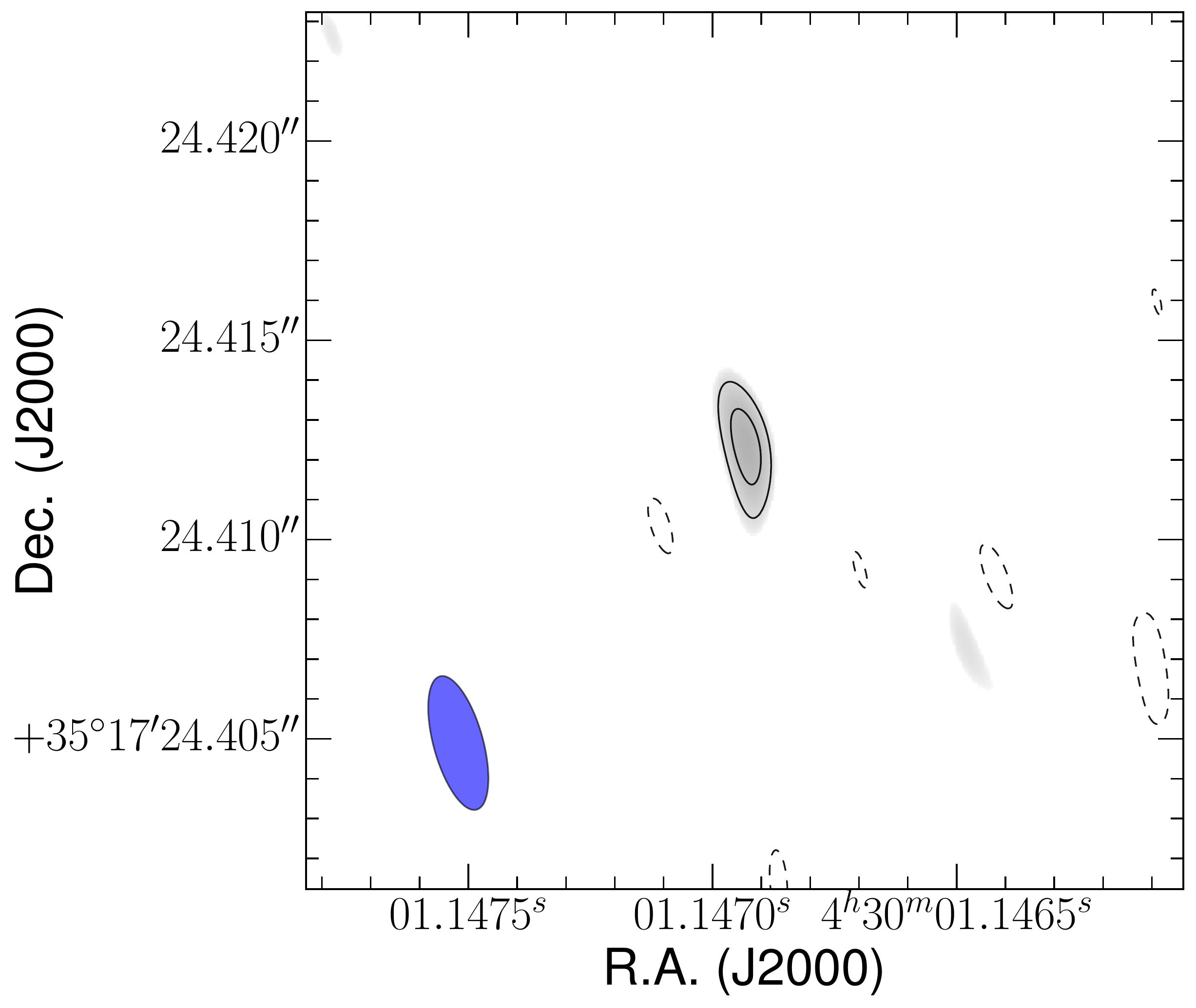}
  \put(-72,160){2017.06.17}&
  \includegraphics[height=0.35\textwidth,trim= 0 0 0 0, clip]{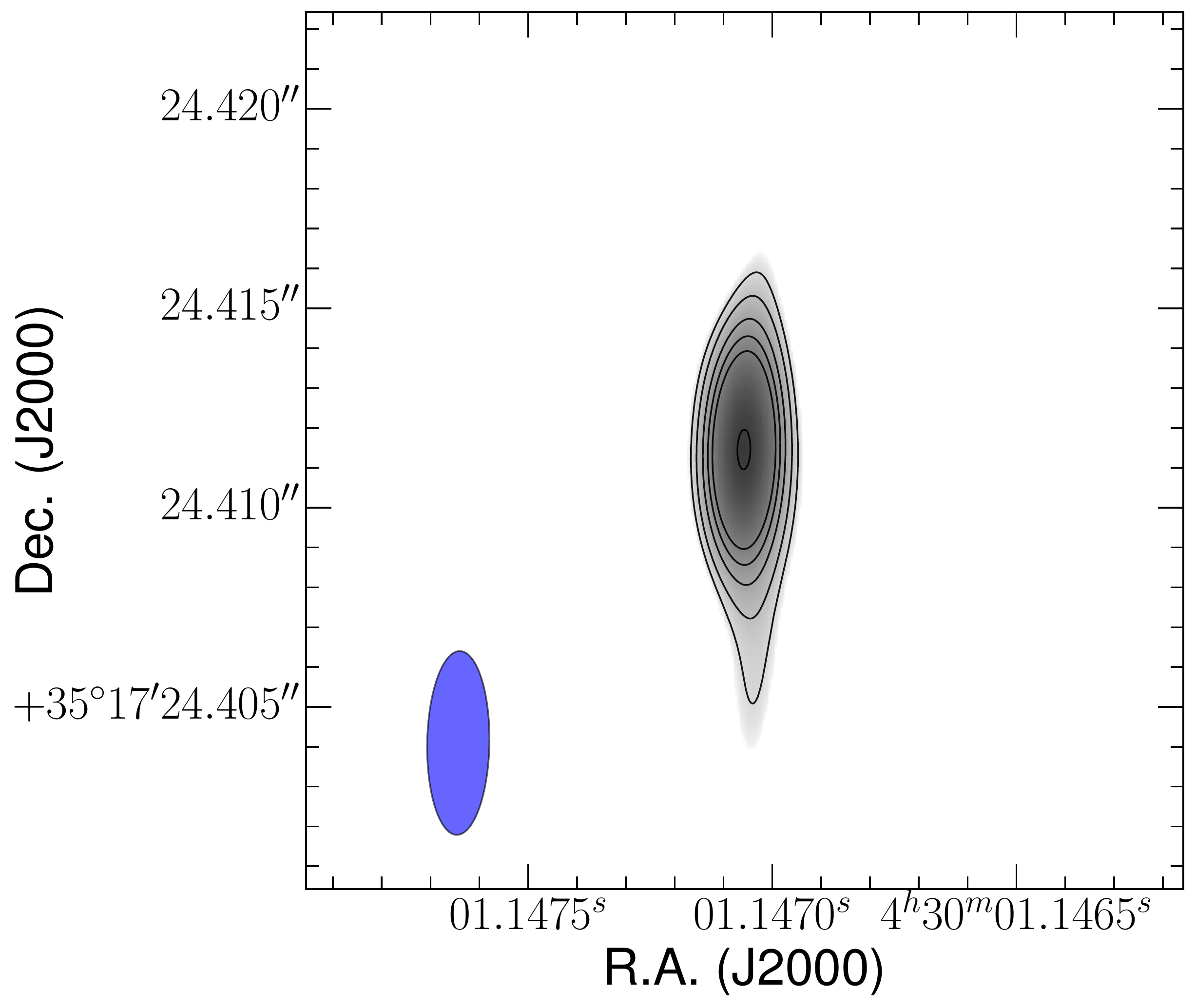}
  \put(-72,160){2017.09.11}
\end{tabular}
\caption{\lkhs\ as detected in each epoch. The contour levels
are as in Figure~\ref{fig:so}. The noise levels and the size of the synthesized
primary beams are listed in Table~\ref{tab:hri}. The latter are displayed as filled 
blue ellipses in the bottom left corner of each image.}
   \label{fig:T}
\end{figure*}

The observations of the targets were recorded as part of cycles with two minutes 
spent on-source and one minute spent on the main phase calibrator, J0429+3319. 
To improve the quality of the phase calibration, we also observed every 30 minutes 
the secondary calibrators J0443+3441, J0414+3418, and J0418+3801. 
Additionally, about two dozen ICRF quasars distributed over the entire visible sky were 
observed  during the observations (conforming the so-called {\em geodetic blocks});
those are used to improve tropospheric calibration \citep[e.g.,][]{reid2004}. 
These geodetic blocks were observed at the beginning
and at the end of each epoch. The observation lengths of each epoch were 3.0  
and 2.5 hours for the projects BD165 and BD207, respectively.

The data were edited and calibrated using the Astronomical 
Image Processing System  \citep[AIPS;][]{Greisen_2003}. 
The basic data reduction followed the standard VLBA procedure
for phase-referenced observations, including 
the multi-calibrator schemes\footnote{The phase transfer from 
the main calibrator to the secondary calibrator J0414+3418 did 
not work properly, so the latter source was excluded from the 
multi-calibrator correction.} and the tropospheric and clock
corrections obtained from the geodetic blocks
\citep[see][for a detailed description of these calibration 
steps]{loinard2007ttau,torres2007,dzib2010}.
After calibration, the visibilities were first imaged with a 
pixel size of 100 $\mu$as using a natural weighting 
scheme (ROBUST = 5 in AIPS) and covering an area of $\sim$1 
square arcsecond. As this scheme provides the best 
possible noise level, we used these images to search for source 
detections. When a detection was obtained, we 
constructed new images, around the source, with a weighting scheme intermediate between natural and uniform 
(ROBUST = 0) using a pixel size of 50 $\mu$as. In these last images we lost some sensitivity, 
but gained some angular resolution, enabling a slightly better determination of the source 
positions at each epoch. These images were then also corrected for the response of the primary beam.
The r.m.s.\ noise levels, $\sigma_{\rm noise}$, in the final images were 
21~--~36~$\mu$Jy~beam$^{-1}$. 
The parameters of the images obtained at individual epochs are given in Table \ref{tab:hri}. From these images, the 
source position, {flux, and deconvolved size were} determined by using a two-dimensional fitting procedure (task JMFIT in AIPS).

\begin{figure}[!ht]
   \centering
  \includegraphics[height=0.7\textwidth,trim= 0 25 15 50, clip]{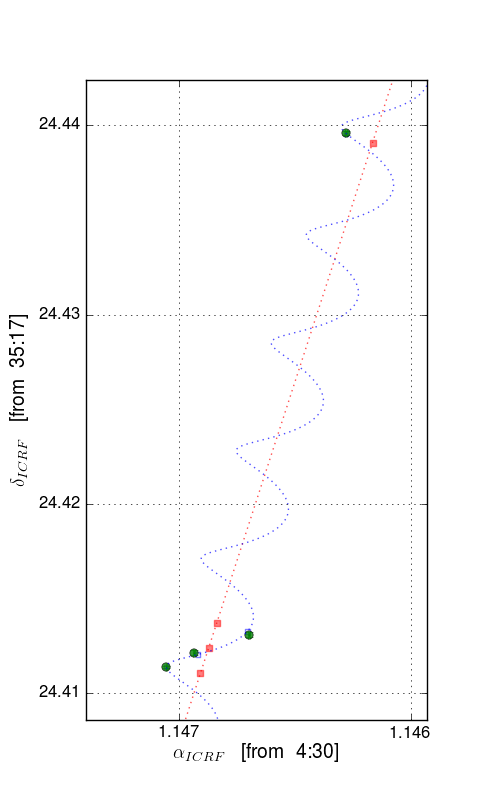}
   \caption{Measured positions (green circles) and best fit  to the 
   movement (dotted blue line) of \lkhs. Blue squares indicate the expected 
   position from the best fit model. The red dotted line is the model
   after subtracting the reflex movement of the trigonometric parallax. {
   The red squares indicate the position of the source at the observed epochs
   expected from the model after correction for reflex parallax motion.} }
   \label{fig:pi}
\end{figure}

\section{Results}


Three of the target sources were detected in our observations (see 
Figures~\ref{fig:so} and \ref{fig:T}, and Table~\ref{tab:targets}). 
\lkh\ VLA 043001.15+351724.6 and  \lkh\ VLA 043017.90+351510.0 were detected as 
single compact radio sources in all four epochs. \lkh\ VLA 043019.15+351745.6, on the contrary, 
{ had a more complex morphology}. In the first epoch a single source was detected, 
however in the last three epochs we detected two sources in each image (e.g., 
bottom of Figure~\ref{fig:so}). This multi-epoch detection of two radio sources
indicates that \lkhc\ is a possible tight binary system, and it will be interesting
for further study. 

The position of the source related to \lkh\ VLA 043017.90+351510.0
did not significantly change between the different epochs suggesting 
that it is a background object and not a member of the \lkh\ cluster. 
The complexity of \lkhc\ and the low number of 
detections make it difficult to perform an accurate astrometric analysis. 
This system will be further analyzed in a future paper when more observations are collected.
The young star \lkh\ VLA 043001.15+351724.4 was well detected in four epochs
and we will focus our astrometric analysis on it. {In the three
first detected epochs, JMFIT cannot deconvolve LkHa 101 VLA J043001.15+351724.6 to a finite size. 
The fourth epoch is also consistent with a point source, although it would also be consistent 
with a deconvolved size up to $0\rlap{.}''0022\times0\rlap{.}''0004$; 
P.A.=$33^{\circ}$.  Thus the target at this epoch might be 
marginally resolved (on account of the unresolved nature of the source at the other epochs, we 
consider that it is more likely that the data contain remaining phase errors).} 
The images of its radio emission at all epochs are shown in Figure~\ref{fig:T}.

\subsection{Astrometry}

The displacement of \lkh\ VLA 043001.15+351724.4 on the plane of the sky can
be modeled as a combination of a trigonometric 
parallax ($\varpi$) and linear proper motions ($\mu$)  \citep[e.g.][]{loinard2007ttau}. The 
fluxes and measured equatorial positions are presented in Table~\ref{tab:hri}. The barycentric 
coordinates of the Earth appropriate for each observation were calculated using the NOVAS routines 
distributed by the US Naval Observatory. The reference
epoch was taken at JD 2457108.94 $\equiv$ J2015.24, the mean epoch of the observations. The best
fit to the data assuming a uniform proper motion (Figure \ref{fig:pi}) yields the following 
astrometric elements:

\begin{ceqn}
\begin{align*}
\alpha_{J2015.24}              &=  04^{{\rm h}}30^{{\rm m}}01\rlap.{^{\rm s}}146538 \pm 0\rlap.{^{\rm s}}000009\\
\delta_{J2015.24}              &=  35^{\circ}17^{'}24\rlap.{''}4251 \pm 0\rlap.{''}0001\\
\mu_\alpha \cos{\delta}    &=  1.86 \pm 0.04\ {\rm mas\ yr}^{-1}\\
\mu_\delta                        &=  -5.70 \pm 0.05\ {\rm mas\ yr}^{-1}\\
\varpi                                &=  1.87 \pm 0.10\ {\rm mas}.
\end{align*}
\end{ceqn}

\noindent This parallax corresponds to a distance of $d$~=~$535\pm29$~pc.
The post-fit r.m.s. values are 0.15 and 0.16 mas in right ascension and
declination, respectively. Systematic errors of 0.10 and 0.13 mas 
(in right ascension and declination, respectively), were added in quadrature 
to the uncertainties delivered by JMFIT to obtain a reduced $\chi^2=1$. 

\section{Discussion and Conclusion}

The measurement of a trigonometric parallax is independent of any assumption 
of the properties 
of the star, since it is a purely geometric method. Consequently, this is a direct 
determination of distances. 

The young star \lkh\ VLA J043001.15+351724.6 is a {\it bona fide} member 
of the \lkh\ cluster \citep{herbig2004,osten2009}. Therefore, its distance 
gives us an accurate approach on the distance to this cluster. The angular size 
of the cluster is $\sim8'$, corresponding to a physical size of 1.25 pc at 
the distance of 535 pc. Because this size is much smaller than our distance error 
we can safely assume that the distance to the cluster, including the \lkh\ star, 
its most massive member, is also $535\pm29$~pc. This result confirms the suggestion 
by \cite{andrews2008} that the distance to the \lkh\ cluster ranges between 500
to 700 pc, and whom also favored a distance of 550 pc. Given this range of values, our result 
has reduced the uncertainty on the distance to the \lkh\ cluster by a factor of three.
Considering that this is the first direct measurement of a distance to one of the 
star members of the \lkh\ cluster, our result is also the most well founded distance to the 
cluster until now.

\acknowledgements
G.-N.O.L acknowledges support from the Alexander von Humboldt 
Foundation in the form of a Humboldt Fellowship. 
L.L. and L.R. acknowledges the financial support of DGAPA, UNAM (project IN112417), and CONACyT, M\'exico. S.-N.X.M. acknowledges IMPRS for a Ph.D. research scholarship.
The Long Baseline Observatory is a facility of the National Science Foundation operated under cooperative agreement by Associated Universities, Inc.


\software{ AIPS (Greisen 2003).}

\bibliographystyle{aa}
\bibliography{references}


\end{document}